\documentclass{ws-ijmpd}
\usepackage[super,compress]{cite}
\usepackage{epsfig,subfigure,latexsym}

\usepackage{amssymb,amsmath}
\usepackage[utf8]{inputenc}
\usepackage{amsfonts}
\usepackage{graphicx}
\usepackage[normalem]{ulem}

\usepackage{multirow}

\newcommand {\bea}{\begin{eqnarray}}
\newcommand {\eea}{\end{eqnarray}}
\newcommand {\be}{\begin{equation}}
\newcommand {\ee}{\end{equation}}

\begin{document}
\def\Journal#1#2#3#4{{\it #1} {\bf #2}  (#4) #3 }
\def\RPP{{Rep. Prog. Phys}}
\def\PRC{{Phys. Rev. C}}
\def\PRD{{Phys. Rev. D}}
\def\ZPA{{Z. Phys. A}}
\def\NPA{{Nucl. Phys. A}} 
\def\JPG{{J. Phys. G }}
\def\PRL{{Phys. Rev. Lett}}
\def\PR{{Phys. Rep.}}
\def\PLB{{Phys. Lett. B}}
\def\AP{{Ann. Phys (N.Y.)}}
\def\EPJA{{Eur. Phys. J. A}}
\def\NP{{Nucl. Phys}}  
\def\RMP{{Rev. Mod. Phys}}
\def\IJMPE{{Int. J. Mod. Phys. E}}
\def\IJMPA{{Int. J. Mod. Phys. A}}
\def\AJ{{Astrophys. J}}
\def\AJL{{Astrophys. J. Lett}}
\def\AA{{Astron. Astrophys}}
\def\ARAA{{Annu. Rev. Astron. Astrophys}}
\def\MPLA{{Mod. Phys. Lett. A}}
\def\ARNPS{{Annu. Rev. Nuc. Part. Sci}}
\def\LRR{{Living. Rev. Relativ}}
\def\APP{{Astropart. Phys}}
\def\CQG{{Class. Quant. Grav}}

\markboth{A. B. Wahidin, A. Rahmansyah, and A. Sulaksono }
{Effect of Scalar Boson on Fermionic Dark Stars}

%
\catchline{}{}{}{}{}
%

\title{Effect of Scalar Boson on Fermionic Dark Stars}
\author{A. B. Wahidin, A. Rahmansyah, and A. Sulaksono}
\address{Departemen Fisika, FMIPA, Universitas Indonesia, Depok 16424, Indonesia.}
\maketitle
\begin{history}
\received{Day Month Year}
\revised{Day Month Year}
\end{history}

\begin{abstract}
The role of scalar boson exchange as a mediator of the fermionic dark particle interaction and the mass of dark particle on the bulk properties of fermionic dark stars including their moment of inertia and tidal deformability are studied. We have found that the role of the attractive nature of the scalar boson exchange and the fermionic dark particle mass can control the stiffness of the fermionic dark star equation of state. By increasing the strength of scalar boson coupling, the fermionic dark star becomes more compact. As a consequence, if scalar boson exchange contribution is included the compactness of a dark star can exceed $C$=0.22. We also compare the fermionic dark stars moment of inertia and tidal deformability to those of neutron stars (with and without hyperons in neutron star core) predicted by relativistic mean field model. It is evidence that the properties of both types of stars are quite different. We also have found that the universal I-Love relation in fermionic dark stars is not affected by scalar boson exchange contribution and the fermionic dark particle mass. Possible observations of fermionic dark stars are also discussed.
\end{abstract} 

\ccode{PACS numbers:95.35.+d, 04.40 Dg}


 \section{INTRODUCTION}
\label{sec_intro}

The existence evidences of dark matter (DM) come from a variety of astrophysical observation due to gravitational behavior of galaxies and clusters. However, because a non-luminous nature of DM, it is difficult to detect DM particles in laboratories, either in terrestrial settings or in satellites.  Therefore, despite many experimental attempts performed  up to now, there is no compelling evidence that DM particles have been observed in laboratories. We have still general guidance information of the properties of DM particles such as they could be bosons or fermions  as well as they are certainly not baryonic and carry no electric or color charges, they are also not composed of standard model (SM) particles, and they do not interact electromagnetically, they should be mostly cold and/or warm and they  interact with SM particles only through gravity.  For review on DM properties see e.g., in Refs.~\cite{PL2017,Olive2003,Munoz2004,Young2017}. Firm and precise information of the properties of DM particles  is still lacking. This information  is very important to provide more understanding about our universe.

The structure, formation, and evolution of the universe from stars to galaxy clusters can be best described  by using $\Lambda$ cold dark matter ($\Lambda$CDM) model. In  $\Lambda$CDM model, the universe consists of 95 $\%$ DM particles and only small portion of particles predicted by the SM. The DM particles in $\Lambda$CDM model are assumed to be cold and collision-less (CCDM)~\cite{Young2017,Kouvaris:2015rea,Maselli:2017vfi}.  However, it is reported recently that the CCDM paradigm is not too compatible with some astrophysical observations.  The corresponding issues are the flatness of DM density profile at the core of dwarf galaxies, the incompatibility of the number of satellite galaxies between prediction by numerical simulations of CCDM and the observed ones, as well as the unobserved massive dwarf galaxies predicted by the CCDM. The discussions about these issues and the corresponding solutions can be found, e.g., in Refs.~\cite{Young2017,Kouvaris:2015rea,Maselli:2017vfi} and the references therein. The idea that DM particles have self-interaction can be considered as one of the compelling solutions for these issues~\cite{Kouvaris:2015rea,Maselli:2017vfi,SS2000}. The interesting consequence of the  DM particles have self-interaction is, the stable but non-luminous compact objects called dark stars (stars composed by DM particles)  can be formed.  Note that DM self-interactions have also been thoroughly studied in various contexts (see details in Ref.~\cite{Maselli:2017vfi} and the references therein). Although depending on the spin of DM particles and the self-interaction forms of DM particles, the appropriate range for the strength of the self-interaction~\cite{Kouvaris:2015rea,Maselli:2017vfi} which sufficient to resolve the CCDM problems, are constrained if we look into the total cross-section $\sigma$ of DM self-interaction. The acceptable ratio of $\sigma$ to dark particle mass $M_{\chi}$ range is  known between (see e.g., in Refs.~\cite{Kouvaris:2015rea,Maselli:2017vfi})
\begin{eqnarray}
  0.1 ~\frac{\rm cm}{\rm g} < \frac{\rm \sigma}{M_{\chi}} < 10 ~\frac{\rm cm}{\rm g}.
  \label{constraint}
\end{eqnarray}
$\frac{\rm \sigma}{M_{\chi}}$ constraint in Eq. ~(\ref{constraint}) is not sufficient to constrain the DM particle mass and the corresponding interactions, simultaneously. However, these quantities can be deduced from dark stars properties if they can be observed. 

The dark stars can not be observed by standard electromagnetic probes. However, in principle dark stars can be observed by gravity probes such as GW signal observations. GW signals~\cite{26,Aasi2015,Acernese2015,Aso2013} from binary coalescence could provide valuable information on the compact objects internal structure.  Up to now, only binary NSs and black holes (BHs) are the known compact objects which are the corresponding GW signals already fully analyzed. It is reported that observation of GW signals which are emitted during neutron star (NS) binary coalescence through the tidal deformability provide very stringent constraint of the allowed nuclear equation of state (EOS) ~\cite{21,22,23,24,Bharat Kumar:2016}. We pointed out here that  tidal deformability extracted from future measurement of GW signal from dark stars binary  can be used also to deduced the allowed EOS of DM. Therefore, how to distinguish dark stars with other compact objects through observables related to GW is our first concern in this work because it may provide important information for the experimentalists in GW detector facilities to detect dark star binary.  We need also to note  that possible GW signals from several possible exotics objects such as boson stars, gravstars, quark stars, and axion stars have been also recently investigated theoretically. They reported that the corresponding GW signals could/could not mimic BHs or NSs (see detail discussions in Ref.~\cite{Sennett2017} and the references therein).

Fermionic DM as a candidate of self-interacting DM is interesting because not only it relates theories beyond standard model to that of DM but also links the baryon-genesis and dark-genesis~(see details in Ref.~\cite{Kouvaris:2015rea} and the references therein for details). In previous study~\cite{Kouvaris:2015rea}, the possibility of self interacting fermionic DM that can solve problems of the CCDM paradigm and can form compact stable objects has been studied. In their work, it is assumed that the fermionic self-interactions can be expressed by non-relativistic Yukawa-type potentials while the dark particle kinetic term is still retain relativistic. The interaction can be either attractive (mediated by a scalar boson exchange) or repulsive (mediated by a vector boson exchange). In their formalism the vector and the scalar potentials have similar form but the only difference is their signs. They have found that in some situations the relativistic effects are important. Furthermore, they obtained the upper mass limit for these compact objects and studied the stability of these objects. They also observed that the fermionic DM stars can rotate faster than NSs. Therefore, it is claimed that any pulsars rotate below a millisecond could be a candidate for a fermionic DM star (or DS for short). Further studied has been done by Masseli $\it et. al$~\cite{Maselli:2017vfi}. They studied the bulk properties of slow rotating DSs including the moment of inertia, tidal deformability and quadrupole moments using the fermionic and bosonic equation of states (EOSs). They using the same model as the one used in~\cite{Kouvaris:2015rea} for fermionic DM case but they considered only the non-relativistic repulsive Yukawa-type interaction and neglected the contribution of  corresponding attractive Yukawa-type interaction  because the contribution is irrelevant in the frame of the used model.  They obtained that the $\rm I-Love-Q$ universal relations are fulfilled also by DSs. They also have proved that the DSs are not compact enough to mimic a black hole in general relativity. Therefore, the gravitational wave interferometers can distinguish both events. Furthermore, they have also shown the compactness of the DSs never exceed the threshold $C\approx $ 0.22. This genuine fact is evidenced not only for bosonic but also for fermionic DM cases. Our second concern here is to check whether if we take care properly the relativistic form not only the kinetic energy but also the scalar and vector potential terms of the model, the conclusions obtained by ~\cite{Kouvaris:2015rea,Maselli:2017vfi} are changed or not. This will be done because we have learned from NS case, that proper relativistic treatment of the interaction potentials are crucial to obtain the correct properties of NS.

In addtion, we need to point out that if DS has only attractive Yukawa-type self-interaction (scalar boson exchange), in some cases to the formation of destructive black holes in the interior of old NSs can be manifested. Therefore, additional constraints are also needed (see the discussions in Ref.~\cite{Kouvaris2012} and the references therein). Furthermore, it is reported that if the fermionic DM interact each other through a scalar boson exchange, huge and stable bound states of DM object called ``nugget'' can be formed. These objects can saturate at a particular sized depending on the corresponding coupling constant, the mass of scalar boson, and the mass of DM particles~\cite{Gresham2017}. The discussions of the impacts of the existence of these nuggets in the synthesis of early universe have been also discussed in Ref.~\cite{Gresham2017} (see also the references therein). We need to note for complateness that the investigation of compact objects composed by fermionic or bosonic dark matter admixed with the neutron star, white dwarf, and strange star matters have been studied for e.g., in Refs.~\cite{PL2017,PM_JS2016,LT_JS2015,AL_MF2010,LHX2012,SC2009,LCL2011,XJZY2014,GMNRT2013,LCLW2013}.  The possibility that compact stars can be formed by non-interacting fermionic DM has been studied also in Ref.~\cite{NSM2006}.

To this end, we should note that  in $\Lambda$CDM model, the cosmological constant $\Lambda$, is phenomenology constant term added to Einstein equation of general relativity (GR). $\Lambda$ makes this model able to fit nicely all key observations.  However, despite the outstanding successes of GR i.e., passed almost all precision tests in intermediate energy scale, GR has still some quite serious problems at both low and high densities including what we have discussed here i.e., the none of experimentally dark matter evidences and a non zero $\Lambda$  (dark energy) problems (see the discussions of recent progress in modified gravity theories in review papers such as~\cite{Berti_etal2015,Will2009,Psaltis2008,Nojiri2017}. This is the reason  a lot of interest in recent years to study the alternative or modified theories to GR such as scalar theories and their generalizations, f(R) theories, theories whose action contains terms quadratic in the curvature, Lorentz-violation theories, massive gravity theories, theories involving non-dynamical fields as well as non-relativistic theories such as MOND. The one of the challenges of the modified gravity theories is, they should consistently describe the early time inflation and late-time acceleration of the universe without introduction of any other dark component~\cite{Nojiri2017}. It means that the observation of dark star can also challenge the existence of the modified gravity theories.

In this paper, we study compact stars made of fermionic DM (DSs). However, here we focus only on the fermionic DM particles interact with each other through exchange not only vector (repulsive) but also scalar (attractive) bosons, simultaneously. The reasons behind this investigation are, first, the role of scalar boson exchange has rather different behavior at high densities compared to the vector one. We explore the role of scalar boson exchange as the mediator of the DM interaction in DSs by studying the bulk properties of slow rotating DSs including the moment of inertia and tidal deformability, and second, now is already the time to discuss the possible observations for DSs through GWs related observables.

This paper is organized as follows: In Sec.~\ref{sec_eos}, we discuss the role of scalar boson exchange on the EOS of the fermionic DM matter at low and high densities. In Sec.~\ref{sec_DS}, we briefly review the formalism used to calculate the bulk properties of slow rotating DSs including the moment of inertia, tidal deformability as well as the corresponding possible observations. The discussion of the obtained results is given also in this section. Finally, the conclusions are given in Sec.~\ref{sec_conclu}.

\begin{figure}
	\centering
	\includegraphics[width=0.8\textwidth]{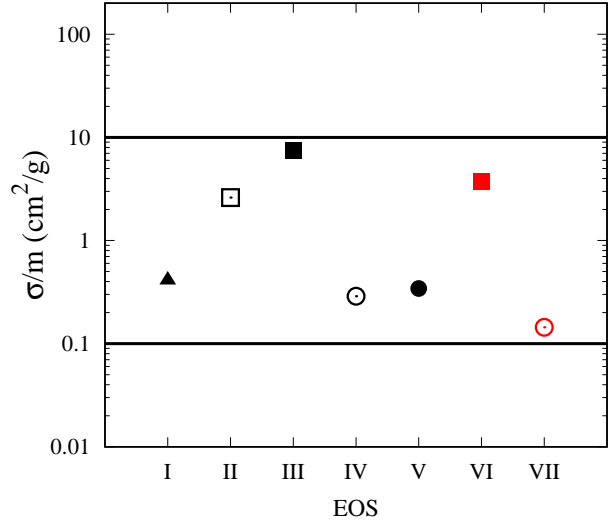}
	\caption{Ratio of total cross section to fermionic DM mass for each EOS which their parameters are defined in Table~\ref{tab:1}. We also include constraint stated in Eq.~(\ref{constraint}).}
	\label{fig:a}
\end{figure}


\section{SCALAR BOSON EFFECT ON EQUATION OF STATE OF SELF-INTERACTING FERMION DARK MATTER}
\label{sec_eos}

In this section, we discuss the features of fermionic DM EOS used in this paper and show the different role of scalar boson exchange as attractive mediator interaction compared to the vector one in EOS of fermionic DM at high densities.

We consider minimal fermionic DM model where the Dirac fermion DM with mass $M_{\chi}$ and field $\psi$ is coupled to scalar $\phi$ and vector $V_{\mu}$ bosons with masses $m_s$ and $m_v$, respectively. Here, the scalar and vector coupling constants are denoted by $g_s$ and  $g_v$. The Lagrangian density of fermionic DM model is expressed as~\cite{Gresham2017,XJZY2014,Diener:2008bj,Glendenning:1997wn}

\bea
{\mathcal{L}}&=&\overline{\psi} [\gamma^{\mu} (i\partial_{\mu}-g_{v} V^\mu)-(M_{\chi}-g_{s} \phi)]\psi\nonumber\\&+&\frac{1}{2}(\partial_{\mu}\phi\partial^{\mu}\phi-m_{s}^2 \phi^2)-\frac{1}{4}\omega_{\mu\nu}\omega^{\mu\nu}+\frac{1}{2}m_{v}^2\omega_{\mu}\omega^{\mu},\nonumber\\
\label{eq:DM}
\eea
where $\omega_{\mu\nu}=\partial^{\mu}V^{\nu}-\partial^{\nu}V^{\mu}$. Using standard relativistic mean field approximation for zero temperature case~\cite{Diener:2008bj,Glendenning:1997wn}, we can obtain the boson fields and the effective mass of fermionic DM easily  as
\begin{eqnarray}
\phi &=& \frac{g_s}{m_s^2}\langle \bar{\psi}\psi \rangle = \frac{g_s}{m_s^2} \frac{\gamma}{2\pi^2} \int_0^{k_F} dk~ \frac{k^2m^*}{\sqrt{k^2+m^{*2}}}\nonumber\\&=&\frac{g_s}{m_s^2}\frac{\gamma}{2\pi^2}\frac{m^{*3}x^3}{2}\chi(x)\nonumber\\
V_0&=&\frac{g_v}{m_v^2}\langle \psi^{\dagger}\psi \rangle = \frac{g_v}{m_v^2} \frac{\gamma}{2\pi^2} \int_0^{k_F}k^2 dk= \frac{g_v}{m_v^2}\frac{\gamma}{3\pi^2}\frac{m^{*3}x^3}{2}\nonumber\\
m^*&=&M_{\chi}-\frac{g_s^2}{m_s^2}\frac{\gamma}{2\pi^2} \frac{m^{*3}x^3}{2}\chi(x),
\end{eqnarray}
where $k_F$ is Fermi momentum and $\gamma$=2 is spin degeneracy factor, while $x = k_F/m^*$ and 
\begin{eqnarray}
\chi(x)\equiv\frac{1}{x^3}\left[x\sqrt{x^2+1} - \mathrm{ln}\left( x + \sqrt{x^2+1}\right)\right].
\label{chi}
\end{eqnarray}
Note that only time component of $V^{\nu}$ ($V_0$) is survived. It can be checked also that for the limit $x \rightarrow 0$, $\chi(x) \rightarrow 2/3$. From energy-momentum tensor of fermionic DM model, we can also obtain the matter energy density $\epsilon$ as
\begin{eqnarray}
\epsilon &=& \tfrac{1}{2} m_s^2\phi^2 + \tfrac{1}{2} m_v^2V_0^2 + \frac{\gamma}{2\pi^2} \int_0^{k_F} dk~ k^2\sqrt{k^2+m^{*2}} \nonumber\\
&=& \frac{\gamma}{2}m^{*4}\xi(x)+\left[ \frac{\gamma^2}{18\pi^3}m^{*6}x^6 \right]\left[ \frac{\alpha_v}{m_v^2}+\frac{9}{4}\frac{\alpha_s}{m_s^2}(\chi(x))^2 \right],\nonumber\\
\label{eq:eps1}
\end{eqnarray}
and pressure $P$ as
\begin{eqnarray}
P &=& -\frac{1}{2} m_s^2\phi^2 + \frac{1}{2} m_v^2V_0^2 +\frac{1}{3} \left( \frac{\gamma}{2\pi^2} \int_0^{k_F} dk~ \frac{k^4}{\sqrt{k^2+m^{*2}}}\right) \nonumber\\
&=& \frac{\gamma}{2}m^{*4}\Psi(x)+\left[ \frac{\gamma^2}{18\pi^3}m^{*6}x^6 \right]\left[ \frac{\alpha_v}{m_v^2}-\frac{9}{4}\frac{\alpha_s}{m_s^2}(\chi(x))^2 \right].\nonumber\\
\label{eq:eps2}
\end{eqnarray}
respectively. Here, we use following definition
\begin{eqnarray}
\xi(x)&\equiv&\frac{1}{8\pi^2}\left[ x\sqrt{x^2+1}\left( 2x^2 +1 \right)  - \mathrm{ln}(x+\sqrt{x^2+1}) \right],\nonumber\\
\Psi(x) &\equiv& \frac{1}{8\pi^2}\left[ x\sqrt{x^2+1}\left( \tfrac{2}{3}x^2 -1 \right)  + \mathrm{ln}(x+\sqrt{x^2+1}) \right],\nonumber\\
\end{eqnarray}
where $\alpha_i ~\equiv ~g_i^2/4\pi$. The form of the expressions in Eqs.~(\ref{eq:eps1})-(\ref{eq:eps2}) is compact and closer to that of Eqs. (11)-(12) in Ref.~\cite{Kouvaris:2015rea,Maselli:2017vfi}.  The non-relativistic limit of Eqs.~(\ref{eq:eps1})-(\ref{eq:eps2}) can be derived easily and they can be expressed as

\begin{eqnarray}
\epsilon &\approx& \frac{\gamma}{2}M_{\chi}^{4} \left[\frac{1}{10}X^5+\frac{1}{10}X^3 \right]+\left[ \frac{\gamma^2}{18\pi^3}M_{\chi}^{6}X^6 \right]\left[ \frac{\alpha_v}{m_v^2}-\frac{\alpha_s}{m_s^2} \right] \nonumber\\
P &\approx& \frac{\gamma}{2} M_{\chi}^{4}\left[\frac{1}{15} X^5\right]+\left[ \frac{\gamma^2}{18\pi^3}M_{\chi}^{6}X^6 \right]\left[ \frac{\alpha_v}{m_v^2}-\frac{\alpha_s}{m_s^2} \right],
\label{eq:epsNR}
\end{eqnarray}
\begin{table}[ph]
\tbl{Parameters of fermionic DM model used in this work to generate the EOSs. We take 2 GeV for fermionic DM mass for EOSs in the upper part of table and 1 GeV in the lower part. EOS II and III have the same pressure as EOS I in low densities, while EOS IV and EOS V have the same energy density as EOS I at low densities. For 2 GeV fermionic DM mass cases, we use the same parameters for vector and scalar couplings as the ones used for EOS III and EOS IV. }
{\begin{tabular}{c|c|c|c|c}
\hline
\multicolumn{5}{c}{$M_{\chi}$ = 2 GeV}\\
\hline
EOS  &$\alpha_v$ & $\alpha_s$ & $\bar{\alpha}_p=\alpha_v-\alpha_s$&  $\bar{\alpha}_{\epsilon}=\alpha_v+\alpha_s$  \\ \hline
I& $10^{-3}$ & $0$ & $10^{-3}$&$10^{-3}$\\
II& $2\times10^{-3}$ & $1\times10^{-3}$ & $10^{-3}$&-\\
III& $3\times10^{-3}$ & $2\times10^{-3}$ & $10^{-3}$&-\\ 
IV& $0.7\times10^{-3}$ & $0.3\times10^{-3}$ &-& $10^{-3}$\\
V& $0.5\times10^{-3}$ & $0.5\times10^{-3}$ &-&$10^{-3}$\\ 
\hline
\multicolumn{5}{c}{$M_{\chi}$ = 1 GeV}\\
\hline
VI& $3\times10^{-3}$ & $2\times10^{-3}$ & $10^{-3}$&-\\ 
VII& $0.7\times10^{-3}$ & $0.3\times10^{-3}$ &-& $10^{-3}$\\
\hline
\end{tabular}
\label{tab:1}}
\end{table}
where $X= k_F/M_{\chi}$.  Therefore, it obvious that in low density or non-relativistic limit, the potentials due to the exchange of scalar and vector bosons behaves exactly the same as a Yukawa-type potential with different coupling and range. The difference is only the scalar boson exchange provides attractive interaction and the vector one provides repulsive interaction.  Furthermore, It can be seen from Eqs.~(\ref{eq:eps1})-(\ref{eq:eps2}) that the scalar boson exchange term enters into $\epsilon$ and $P$ differently especially at high densities. The behavior of scalar boson exchange is not only different to that of vector boson in the interacting part of $\epsilon$ and $P$ (with a different sign) but it also enhances the kinetic part of  $\epsilon$ and $P$ through its contribution in effective mass $m^*$.  Therefore, it seems that the role of scalar boson exchange at high densities (relativistic region) cannot be fully replaced by using only the vector boson with adjusting coupling constant. The impacts of the difference due to the presence of scalar boson exchange at high densities is our main focus in this investigation.

Here, we generate a set of DM EOSs by using Eqs.~(\ref{eq:eps1})-(\ref{eq:eps2}), parameters $M_{\chi}\equiv$ (1-2) GeV, $m_v$ = $m_s$   $\equiv~$10 MeV, and some dark fermion-boson coupling constants combination which are shown in Table~\ref{tab:1}. The total cross section expression can be derived easily and it takes following form: 
\begin{eqnarray}
\sigma_{tot} = \sigma_s+\sigma_v=\frac{M_\chi}{16\pi}\left(\frac{7 g_s^4}{m_s^4}+\frac{3 g_v^4}{m_v^4}\right).
\label{eq:crosssection}
\end{eqnarray} 
As shown in Fig.~\ref{fig:a}, the corresponding parameters in Table~\ref{tab:1} are still compatible with the total cross section constraint at low energies \cite{Kouvaris:2015rea,Maselli:2017vfi}. It can be seen that for fixed $\bar{\alpha}_p$ value the cross-section increases by raising $\alpha_s$ but for the case of fixed $\bar{\alpha}_{\epsilon}$ the cross-section is relative constant by raising $\alpha_s$. Both are evaluated for fixed dark particle mass.

\begin{figure}
	\centering
	\includegraphics[width=0.7\textwidth]{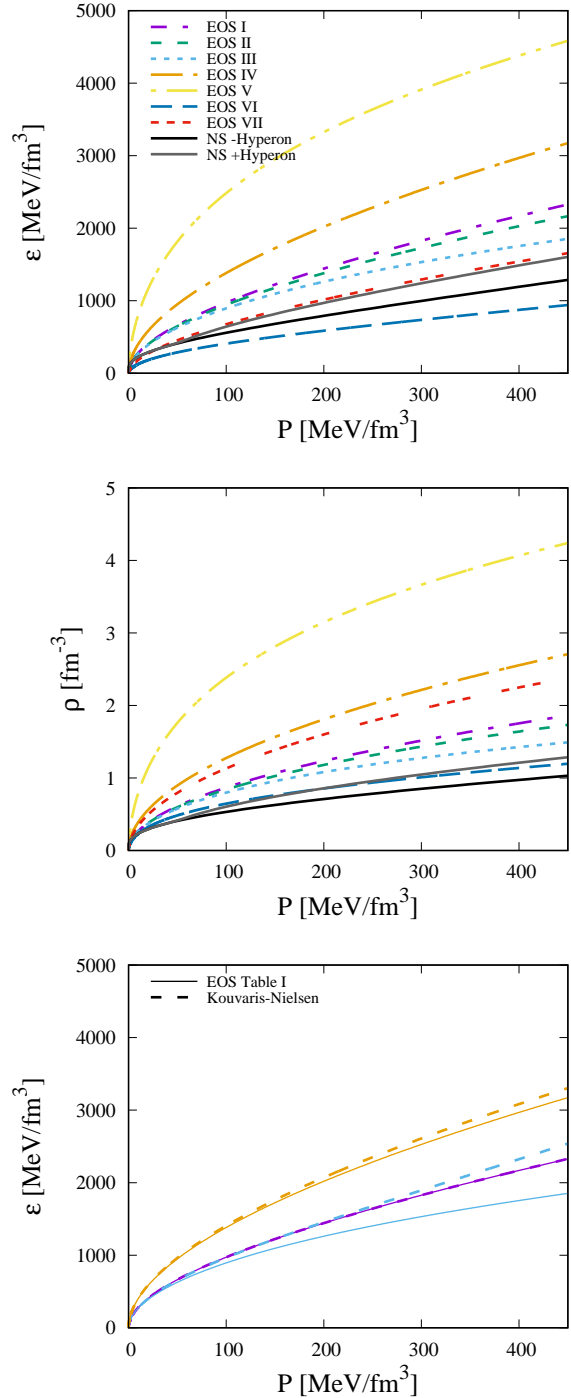}
	\caption{EOSs of fermionic DM model using parameter sets state in Table~\ref{tab:1}, we also include EOS of NS with and without hyperons for comparison (upper panel). Number density as a function of pressure (middle panel). The effect of exact treatment of scalar boson exchange compared to the approximate one used by Kouvaris-Nielsen\cite{Kouvaris:2015rea} formalism for the cases of EOS I, EOS III, and EOS IV (lower panel).}
	\label{fig:1}
\end{figure}

In lower panel of Fig.~\ref{fig:1}, we compare the  EOS as a function of pressure obtained by using Eqs. (\ref{eq:eps1})-(\ref{eq:eps2}) with the ones obtained by using Eqs.(11)-(12) in Ref.~\cite{Kouvaris:2015rea} using exactly the same parameter sets used for EOS I, EOS III, EOS IV. It can be seen that for the case of EOS I (interaction is mediated only by vector boson exchange) the lines coincide, while for the case EOS IV (similar to $\bar{\alpha}_{\epsilon}$ value of EOS I), the stiffness difference is not too significant and the difference appears at high pressures. However, for the case EOS III (similar to $\bar{\alpha}_p$ value of EOS I), the corresponding difference is quite significant as we expected. In general, EOSs predicted by  Eqs.~(\ref{eq:eps1})-(\ref{eq:eps2}) are relative stiffer than those their counterpart in  Eqs. (11)-(12) of Ref.~\cite{Kouvaris:2015rea}.

In upper and middle panels of Fig.~\ref{fig:1}, we compare the  EOSs and number densities as a function of pressure obtained by using EOS I-EOS V for $M_{\chi}=$ 2 GeV and EOS VI and EOS VII for $M_{\chi}=$ 1 GeV. For comparison, we include also the NSs results obtained by using relativistic mean field (RMF) model by employing the BSP parameter set with and without hyperons~\cite{SB2012,QISR2016}. For the case NS with hyperons we use SU(6) as a standard prescription for determining the hyperons coupling constants. The stiffness of BSP parameter set EOS is in between the one of the stiffest RMF EOS (NL3) and the softness RMF EOS (FSU). The contribution of hyperons makes the NS EOS softer. Therefore, the corresponding maximum mass prediction is lower than 2.0  $M_\odot$ if the prescription for determining hyperon coupling constants did not modify or other exotic effects were included in corresponding EOS. It is known as "Hyperon puzzle". See details about the NS EOSs, the compatibility of its predictions with experimental data and the corresponding model parameters in Refs.~\cite{SB2012,QISR2016}.  The corresponding combinations of the scalar and vector coupling constants of EOS I- EOS VII provide the total cross-sections which are still compatible with the constraint in Fig.~\ref{fig:a}. It makes the fermionic DM EOS for $M_{\chi}=$ 2 GeV  softer at high densities compared to the one of NSs with hyperons predicted by BSP parameter set. The role of the scalar field is crucial to control the stiffness of fermionic DM EOS at high densities. Increasing the scalar field contribution using fixed $\bar{\alpha}_{\epsilon}$ tends to make the EOS softer while increasing the strength of scalar field contribution using fixed $\bar{\alpha}_P$ tends to make the EOS stiffer. However, for the case $M_{\chi}=$ 1 GeV (decreasing dark particle mass), the stiffness of the corresponding EOSs are comparable to those of NS EOSs. Concerning with how to distinguish the NS and DS,  the corresponding stiffness similarity of the EOS between the one of DS with $M_{\chi}=$ 1 GeV and the one NS need to be scrutinized. This matter will be discussed in next section.

\begin{figure}
	\centering
	\includegraphics[width=0.7\textwidth]{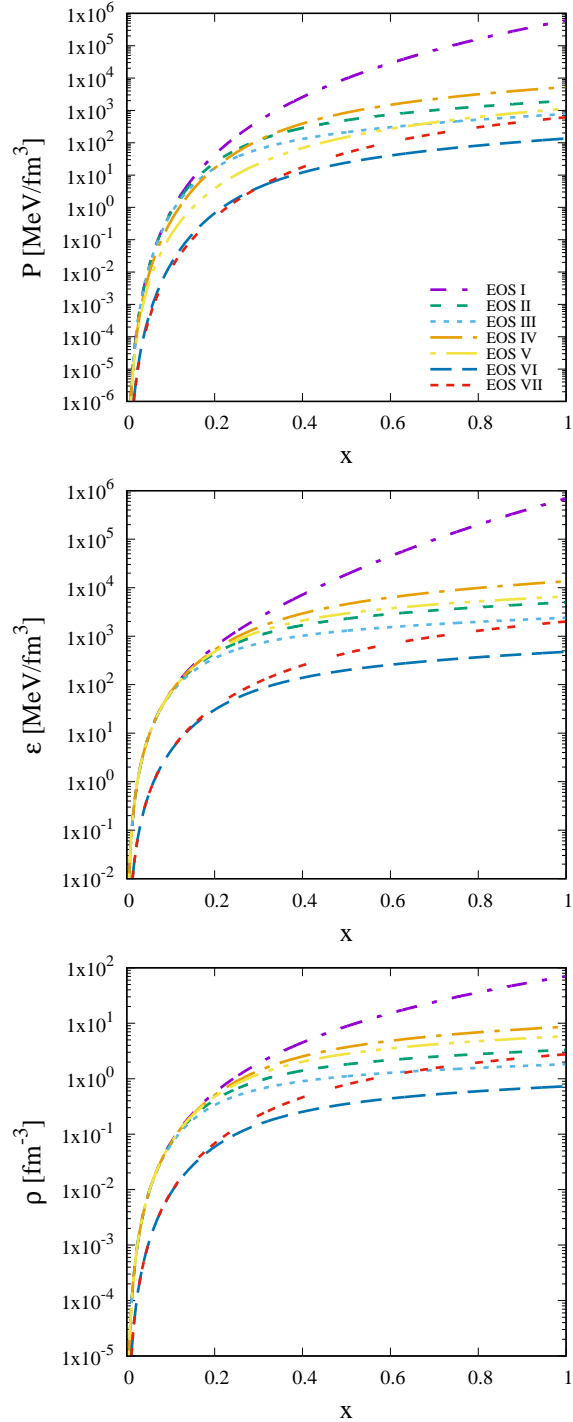}
	\caption{Fermionic DM EOS pressure (upper panel), energy density (middle panel), and number density (lower panel) as a function of relativistic factor $x=k_F/m^*$.}
	\label{fig:2}
\end{figure}

At the upper panel of Fig.~\ref{fig:2}. We can also observe that all fermion DM matter properties such as pressure, energy density and number density are identical at low densities or non-relativistic limit($x<<1$). However, at higher densities or in relativistic regions, it is shown that the behavior of the corresponding EOSs is significantly different from one to another. The effects on energy density and number density can be seen also in middle and lower panels of Fig.~\ref{fig:2}. These stiffness differences of the corresponding EOSs yield significant effects on the properties of DSs. This shows that the properly consider the relativistic effect specially for scalar boson contribution is important for the stiffness of DM EOS. This effect yields impact in the properties of DS. 

\begin{figure}
	\centering
	\includegraphics[width=0.8\textwidth]{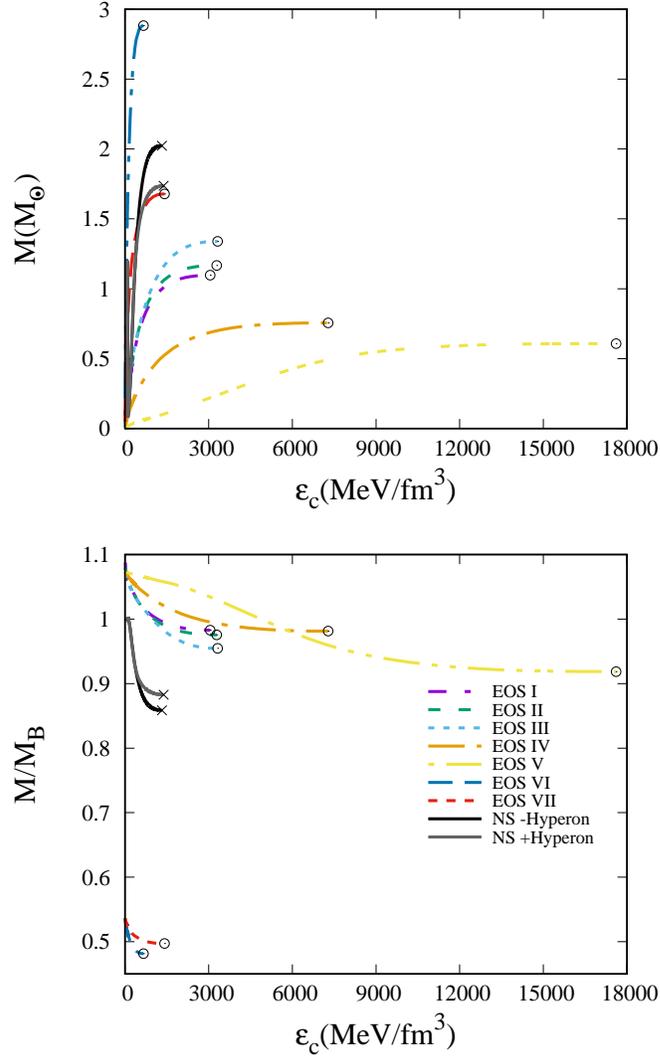}
	\caption{DS (gravitation) mass (upper panel) and the ratio of DS mass to dark particles mass (lower panel) as function of energy density in the center of the fermionic DS.}
	\label{fig:3}
\end{figure}

\begin{figure}
	\centering
	\includegraphics[width=0.65\textwidth]{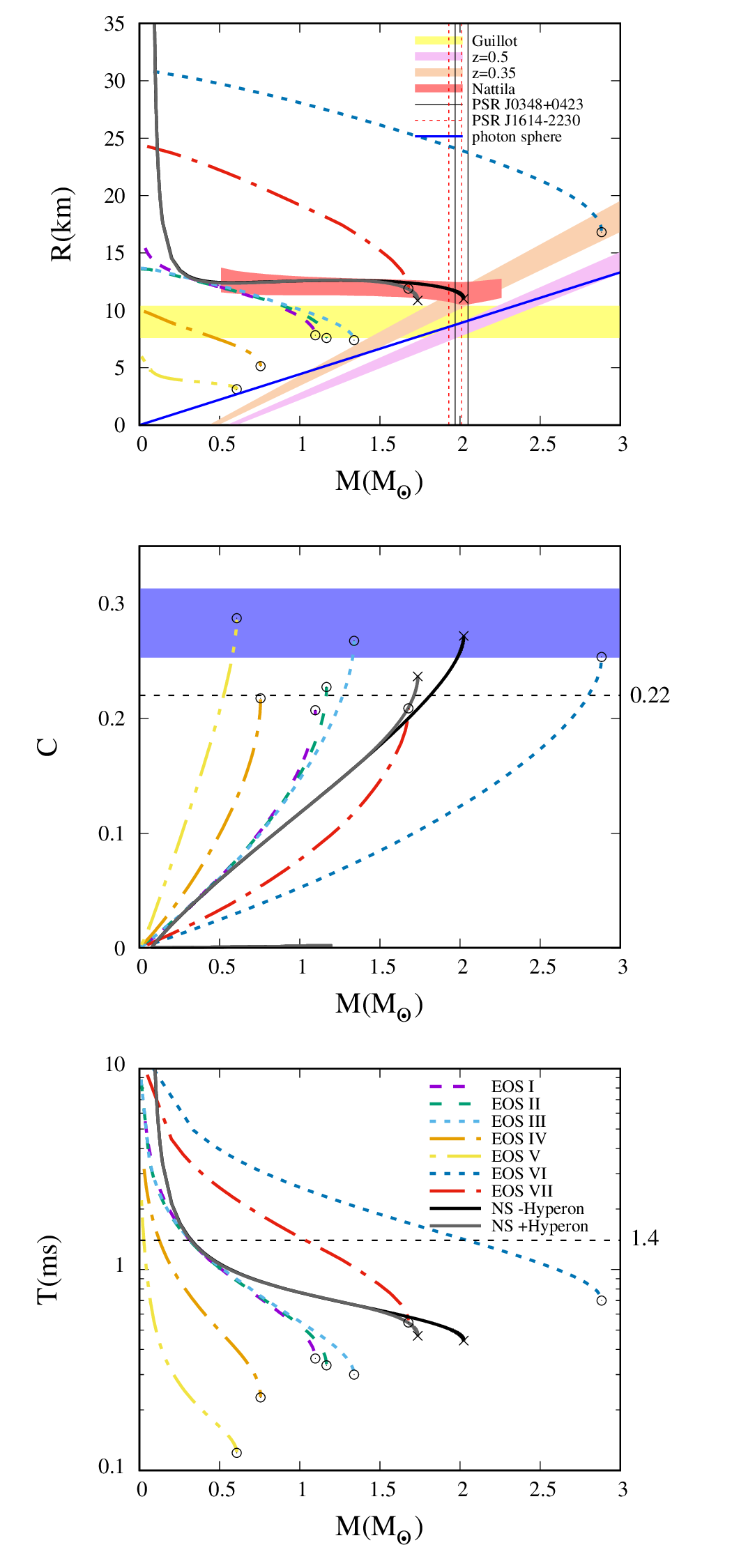}
	\caption{Fermionic DS mass-radius relation obtained by using parameter sets shown in Table~\ref{tab:1} (upper panel). The shaded regions correspond to regions of NS constant surface red-shift z = 0.5 and z = 0.35 with 10\% accuracy. For comparison, we  also show recent observational constraints on NS masses and radii from \cite{Nattila:2015jra}, radius constraint $R_{NS} = 9.1^{+1.3}_{-1.5}$ km (90\%-confidence) from Guillot \cite{Guillot:2013wu}. Compactness $C=GM/R$ as a function of DS mass (middle panel). Diagonal black line is photon-sphere line\cite{28}. We also include NS compactness constraint from \cite{Palenzuela:2015ima} i.e., $0.283 \pm 0.030$ and black dashed line is the bound result taken from Maselli \cite{Maselli:2017vfi}.  Minimum period, according to Keplerian limit $T=2\pi/\Omega_k$ (lower Panel). The dashed horizontal lines represents the fastest known pulsar with $f$=716 Hz.}
	\label{fig:4}
\end{figure}

\section{PROPERTIES OF DARK STARS}
\label{sec_DS}

In this section, we briefly describe the formalism used to calculate macroscopic quantities which characterize the DSs such as mass-radius relation, the moment of inertia, tidal deformation, the universal relation between moment inertia and tidal deformation of DSs, as well as the possible observations and followed by showing the results and presenting the corresponding discussions.

\subsection{Mass-radius relation}
\label{subsec_mass-rad}

We can observe the stability border from the maximum (gravitation) mass of the star or from the minimum ratio of gravitation mass and DM particle mass which is related to binding energy~\cite{Kouvaris:2015rea}. The corresponding plots have been shown in upper and lower panels of Fig.~\ref{fig:3}. It is obvious that stability border of DSs depends significantly on the effect of the strength of scalar boson coupling and DM particle mass. It can be seen that for fixed $\bar{\alpha}_p$ and DM particle mass case, increasing the strength of scalar field contribution makes the maximum mass increases, but the $\epsilon_c$ value in the stability border is not significantly changed. However, for fixed $\bar{\alpha}_{\epsilon}$ case, increasing the strength of scalar field contribution makes the maximum mass decreases and the $\epsilon_c$ value in the stability border depends significantly on the strength of scalar field contribution. In case of EOS V (the softest DM EOS in our cases) for e.g., we obtain the maximum mass around 0.5 $M_\odot$ and the maximum stable center density of fermionic DS even about $\epsilon_c \approx 17610~ \rm MeV/fm^3$. However, because in the $M_{\chi}=$ 2 GeV case, the DS EOSs are softer than those of representative NS EOSs (BSP with and without hyperons), then we obtain smaller maximum mass and the stability border presents in relative larger center energy density than that of our representation NS soft EOS (BSP with hyperons). Note that we can obtain relative larger maximum mass and smaller ratio $M/M_B$ of DSs with $\epsilon_c$ values are similar to those of NSs by decreasing the DM particle mass parameter (see EOS VI and EOS VII). However, this attempt does not significantly change the compactness behavior \cite{Maselli:2017vfi} because by increasing mass, the radius also increases.

In the upper panel of Fig.~\ref{fig:4} we show several observations of NS. Solid black line and dashed red line are the results from precisely measured NSs masses, such as PSR J1614-2230 with mass $M = (1.97 \pm 0.04) M_\odot$\cite{Demorest10} and PSR J0348+0432 with mass $M = (2.01 \pm 0.04) M_\odot$\cite{Antoniadis13}. We also show two shaded regions corresponding to constant surface red-shift defined as
\begin{equation}
\label{eq:zredshift}
z = \left( 1-\frac{2GM}{R} \right)^{-1/2}-1
\end{equation}
with two NS observations reference value,  i.e.,  $z$ = 0.35 and $z$ = 0.5 (see Ref.~\cite{Maselli:2017vfi} and the references therein for details about red-shift constraints). Also we show NS mass-radius constraint from N$\ddot{\rm a}$till$\ddot{\rm a}$~\cite{Nattila:2015jra} and radius constraint from Guillot \cite{Guillot:2013wu}. It can be seen that from this panel that DSs has a very broad range of mass and radius. We can see that all DSs with EOSs parameters from Table.~\ref{tab:1} yield significantly different mass and radius predictions from those of our representative NSs (with and without hyperons). In the case of $M_{\chi}$= 2 GeV, some of the radius predictions are compatible with Guillot NS radius constraint\cite{Guillot:2013wu}. Furthermore, EOS IV and EOS V yield smaller radii than that of the radius constraint range of Ref.\cite{Guillot:2013wu}. However, in the case of $M_{\chi}$= 1 GeV, the corresponding radii can be larger than that of the NS radius constraint of  Ref.\cite{Guillot:2013wu}. It can be seen also except the one of EOS VI, that our mass-radius relations are not compatible with both NS red-shift constraints. The results which are obtained by using EOS I-V and EOS VII, confirm the ones obtained in Ref.~\cite{Maselli:2017vfi}.

The DS compactness as a function of DS mass is shown in the middle panel of Fig.~\ref{fig:4}. We can see that the scalar boson exchange makes DS more compact. This happens because of the attractive nature of scalar boson exchange contribution. Contrary to the results obtained by Maselli $\it ~et ~al.$\cite{Maselli:2017vfi}, except the one of EOS I, DSs can exceed the threshold C $\approx$ 0.22 Even EOS V and EOS III are quite compatible to NS compactness constraint of Palenzuela-Libling\cite{Palenzuela:2015ima}.  This shows that the consequency of properly consider the relativistic effect specially for scalar boson contribution is the compactness of DS can exceed C = 0.22. However, our results are still bellowing the well known maximum theoretical bound for stationary and static NS compactness i.e., $\leq$ 4/9 $\sim$ 0.44. It means that DSs cannot act like black hole mimickers (i.e., a compact object with compactness approaching the limit C $\rightarrow$ 0.5). Therefore, DSs can be distinguished from black holes from the corresponding compactness. This finding is compatible with the one obtained by Maselli $\it et ~al.$\cite{Maselli:2017vfi}.

In addition, it is also recently discussed that gravitational wave echoes at a frequency of about 72 Hz have been produced in the GW 170817 event. The existence of these echoes could be indicated by the mass-radius plots of ultra-compact stars which are crossing the photon-sphere line. The authors of Ref.~\cite{28} has shown that strange stars can emit gravitational wave echoes on the order of tens of kilohertz (see details in Ref.~\cite{28} and the references therein). It is obvious from the upper panel of Fig.~\ref{fig:4} the all DS EOSs used in this work can not emit these echoes. It means that DSs can be distinguished from strange stars in this way.

The bottom panel of Fig.~\ref{fig:4} shows the minimum rotational period derived from the Keplerian limit $T=2\pi/\Omega_k$, where $\Omega_k\approx\sqrt{GM/R^2}$ (mass-shedding limit). The horizontal dashed line corresponds to maximum frequency observed for a spinning NSs, $f=716$Hz or $T\approx1.4$ms. Similar to the finding reported in Ref.\cite{Maselli:2017vfi}, it can be observed that  DSs spin faster than NSs only for the case DM particle with $M_{\chi}$= 2 GeV. However, for EOS VI, DS could be slower spinning than those of NSs. The effect of scalar boson exchange as a mediator of attractive interaction tends to rotate the DS faster than those of NSs only for the case of $M_{\chi} \ge $ 2 GeV.

\subsection{Moment of inertia }
\label{subsec_slowrot}
\begin{figure}
	\centering
	\includegraphics[width=0.8\textwidth]{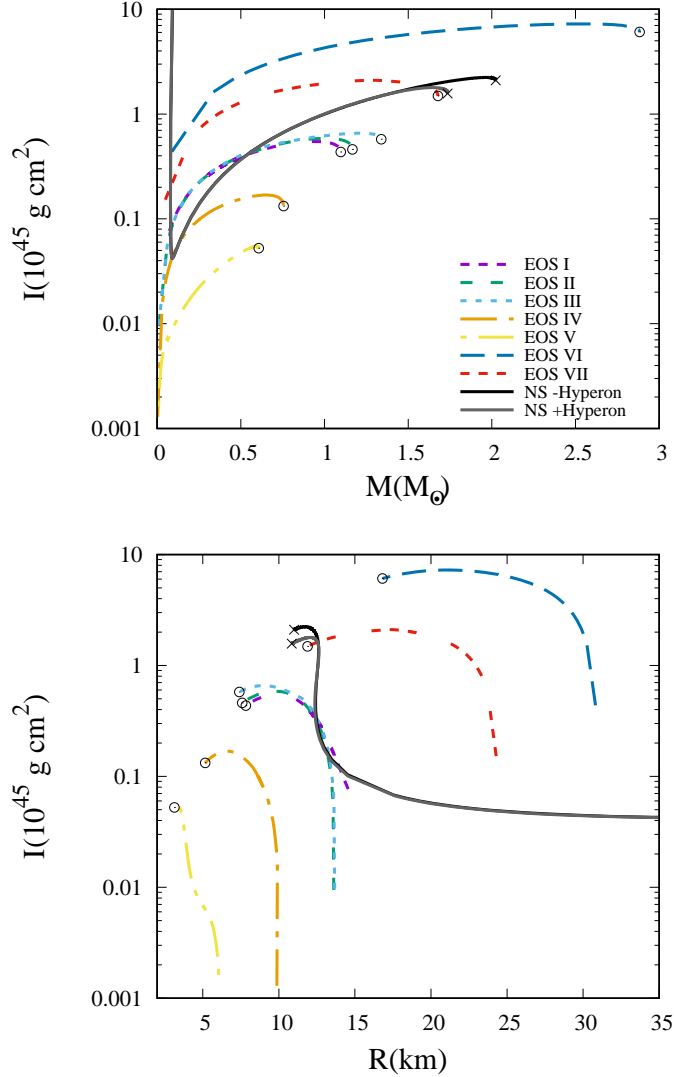}
	\caption{Fermionic DS moment of inertia as a function of  mass (upper panel) and radius (lower panel). }
	\label{fig:5}
\end{figure}

\begin{figure}
	\centering
	\includegraphics[width=0.8\textwidth]{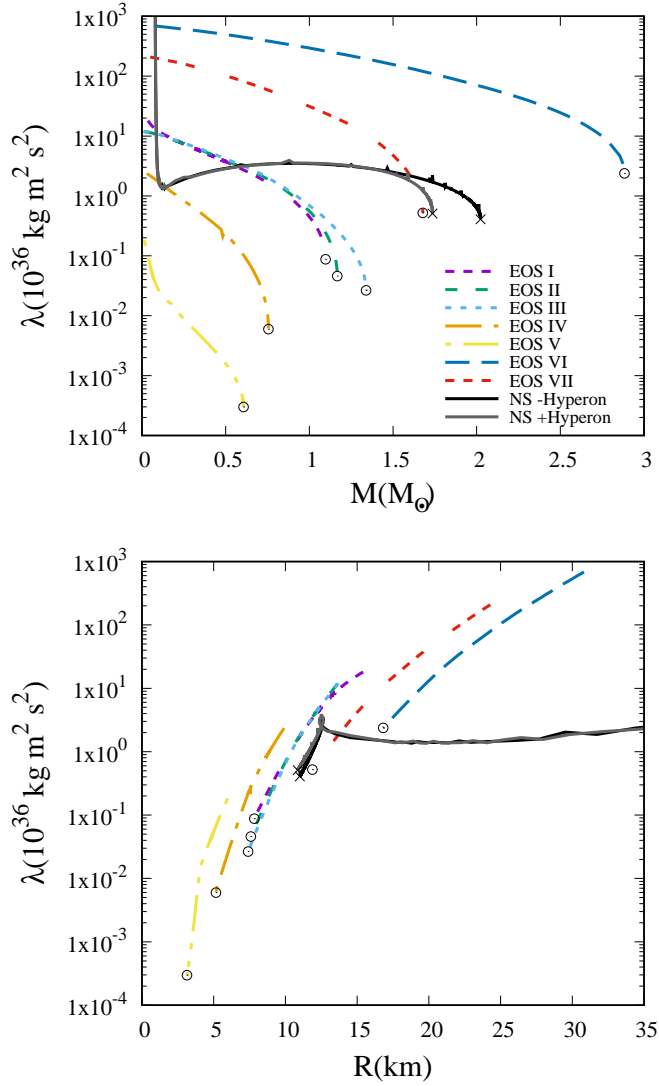}
	\caption{Fermionic DS tidal deformability $\lambda$ as a function mass (upper panel) and radius (lower panel). }
	\label{fig:6}
\end{figure}
To investigate the DS rotating properties, we start with the line element  of slowly rotating compact objects as \cite{IOJ2015,Suparti:2017msx},
\begin{eqnarray}
ds^2 &=& -e^{2\nu}dt^2+e^{2\lambda}dr+r^2(d\theta^2+\sin^2\theta d\phi^2)\nonumber\\&-&2\omega(r)r^2\sin^2\theta dt d\phi,
\label{LEE}
\end{eqnarray}
here $\omega(r)$ accounts for the frame-dragging effect due to rotation. From the 
definition of $\bar{\omega}(r) \equiv \Omega-\omega(r)$, it appears 
that the line element given in Eq.~(\ref{LEE}) is  only correct up 
to order of $\Omega$. The slowly rotating approximation used here means that the  compact object still retains its spherical form, since the centrifugal deformation is considered to be at higher order, i.e.,  order of $\Omega^2$~\cite{IOJ2015}. Solving the Einstein equation by means of Eq.~(\ref{LEE}) and assuming that the fermionic DM in DS is a perfect fluid, one can obtain the Tolman-Oppenhaimer-Volkoff (TOV) equations as
\begin{eqnarray}
\frac{dm}{dt} &=& 4\pi\epsilon r^2 , \\
\frac{dP}{dr} &=& -G\frac{\epsilon m}{r^2}\left( 1 + \frac{P}{\epsilon} \right) \left( 1+\frac{4\pi r^3 P}{m} \right) \left( 1-\frac{2Gm}{r} \right)^{-1} ,\nonumber\\
\label{TOV}
\end{eqnarray}
and the the $\nu$ function in metric fulfills following equation 
\begin{eqnarray}
\frac{d\nu}{dr}	= G\frac{m+4\pi r^3 P}{r(r-2Gm)},
\end{eqnarray}
while $\bar{\omega}$ function  fulfills  second order ordinary differential equation as follow
\begin{eqnarray}
\frac{1}{r^4}\frac{d}{dr}\left[ r^4e^{-\nu}\left( 1-\frac{2Gm}{r} \right)^{1/2} \bar{\omega} \right]\nonumber\\+\frac{4}{r^4}\left[ \frac{d}{dr}e^{-\nu}\left( 1-\frac{2Gm}{r} \right)^{1/2} \right]\bar{\omega} =0.
\label{omega}
\end{eqnarray}
Using the DS EOS $P=P(\epsilon)$ calculated by means of the 
method given in section \ref{sec_eos}, Eqs.~(\ref{TOV})-(\ref{omega}) 
can be numerically integrated by utilizing a standard Runge-Kutta method with boundary values: $P(\approx 0)$=$P_c$, $m(\approx 0)$ $\approx 0$, $P(R)$ $\approx 0$, $m(R)$=$M$. Here $R$ is radius and $M$ is mass of the star. Since at $R$, ${d\bar{\omega}}/{dr} = { 6 G I \Omega}/{R^4}$, the moment of  inertia $I$ can be obtained from Eq.~(\ref{omega}).   

Moment of inertia is a  potential observable which depends more on the stars compactness rather than on the details of the micro-physics properties of the corresponding stars. The bound of moment inertia of compact objects is strongly correlated with the star radius constraint. Moreover, moment inertia affects different astrophysical process such as pulsar glitches and the spin-orbit coupling in the compact binary system (see the corresponding discussion in Ref.\cite{Maselli:2017vfi} and also in the references therein). We have shown in upper and lower panels of Fig.~\ref{fig:5}, the moment of inertia as a function of DS mass and radius.  It can be observed that except the one of EOS VI, the moments of inertia of the DSs are lower than those of our representative NSs. It can be seen also that the different behavior of moments of inertia in low mass DS compared to the ones of NSs is due to the fact that DSs have no crust in their surfaces. Note that for the moment of inertia and radius of NSs measurements are deduced mostly from electromagnetic signature which is absent for DSs. However, due to the fact that moment of inertia has universal relation with tidal deformation and the tidal deformability parameter is expected to be determined from gravitation wave signals, we can also deduce the moment of inertia or radius of compact objects indirectly from the information given by tidal love number. This could be also used to show the existence of DSs. This matter will be discussed in subsection~\ref{subsec_TLN}. Note that recently, it is  reported that the new constraints on radii from tidal deformability of NSs from GW170817 are already studied in Refs.~\cite{27,MGF2013,MWRS2018,CHZ2018,29}.

\subsection{Tidal love number}
\label{subsec_TLN}
The gravitational wave signals which are emitted during binary coalescence, can provide valuable information about the EOS of compact objects. The signals are mostly determined by adiabatic tidal interactions.  Adiabatic tidal interactions are characterized by Love numbers where the dominant contribution $k_2$, associated with quadrupole deformation (see Ref.~\cite{Maselli:2017vfi} and the references therein for further discussions). Furthermore, Love number mainly depends on the compactness of the stars~\cite{Bharat Kumar:2016,TH2010}. Note that the authors of Ref.~\cite{Maselli:2017vfi} have computed the constraints that current gravitation wave interferometers may be able to set from signals emitted binary systems composed of two DSs.

Here we briefly review the essential equations to describe the DS tidal deformability. Detail discussions of the NS tidal deformability can be found in Refs.~\cite{Bharat Kumar:2016,TH2010}. $\lambda$ describes deformability from spherical compact stars when placed in a static external quadrupole tidal field $\epsilon_{ij}$ \cite{Bharat Kumar:2016} as
\begin{eqnarray}
Q_{ij}= -\lambda \mathcal{E}_{ij}~~~~~~~~~~~~~~~\lambda = \frac{2}{3}k_2 R^5, 
\end{eqnarray}
where $k_2$ is a dimensionless quantity called Love number. Its expression can be written as  \cite{Bharat Kumar:2016}
\begin{eqnarray}
k_2 &=& \frac{8}{5}(1-2C)^2 C^5[2C(4(y_2-1)-y_2+2]\nonumber\\&&\Bigg\lbrace  2C(4(y_2+1)C^4+(6y_2-4)C^3 + (26-22y_2)C^2 \nonumber\\
&&+ 3(5y_2-8)C - 3y_2+6)-3(1-2C)^2(2C(y_2-1)\nonumber\\&-&y_2+2)\log\left( \frac{1}{1-2C} \right)\Bigg\rbrace^{-1}.
\end{eqnarray}
Here $C =G M/R$ is compactness parameter while the $y_{2}$ fulfill following first order differential equation  
\begin{equation}
r\frac{dy_l(r)}{dr}+y_l(r)^2+y_l(r)F(r)+r^2Q_l(r)=0,
\end{equation}
here in general $l=2,3,4...$. Note that the explicit form of $F(r)$ is 
\begin{equation}
F(r)= 1 - 4\pi r^2 G (\epsilon-P) \left( 1-2\frac{Gm}{r} \right)^{-1},
\end{equation}
and $Q_l(r)$ is
\begin{eqnarray}
Q_l(r)&=&\left[ 4 \pi r^2 G \left( 5\epsilon+9 P+\frac{\epsilon+P}{\partial P /\partial\epsilon} \right)  - l(l+1) \right]\nonumber\\
&&\left( 1-2\frac{Gm}{r} \right)^{-1}\nonumber\\
&-&4\left( \frac{Gm}{r} \right)^2 \left( 1+\frac{4\pi r^3 P}{M} \right)^2 \left( 1-2\frac{Gm}{r} \right)^{-2}.
\end{eqnarray}
\begin{figure}
	\centering
	\includegraphics[width=0.8\textwidth]{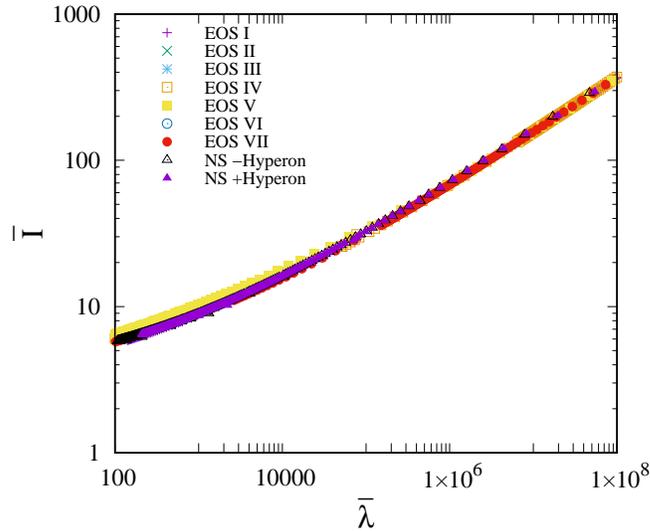}
	\caption{I-Love relation of DSs using parameter sets  which are presented in Table~\ref{tab:1}.}
	\label{fig:7}
\end{figure}
The differential equation of  $y_{2}(r)$ is solved simultaneously with TOV equations. The boundary conditions to solve $y_{2}(r)$  equation are  $r~\approx 0$,  $y_{2}(\approx 0)=2$ and at $R$,   $y_{2}(R)=y_{2}$. 

We have shown in upper and lower panels of Fig.~\ref{fig:6}, the tidal love number $\lambda$  as a function of DS mass and radius, respectively  using EOSs in Table~\ref{tab:1}. It can be observed that $\lambda$ behavior of DSs is relatively different from those of the representative NSs. It can be seen that the $\lambda$ at maximum mass or minimum stable radius depends significantly on the strength of the scalar boson exchange coupling and the mass of DM particles.
By decreasing stiffness of the EOS,  $\lambda$ at maximum mass becomes smaller. We also check for I-Love universal relation by plotting $\bar{I}$ vs $\bar{\lambda}$ where the results are shown in Fig.~\ref{fig:7}. Here, we used $\bar{I}$= $I/M^3$ and  $\bar{\lambda}$=$\lambda/M^5$ relations. It can be seen obviously that this relation is fulfilled independently to the EOS used. These  results are compatible with the ones obtained in Ref.\cite{Maselli:2017vfi}. We will discuss this universal relation further in subsection \ref{subsec_observ}. 

\begin{figure}
	\centering
	\includegraphics[width=0.65\textwidth]{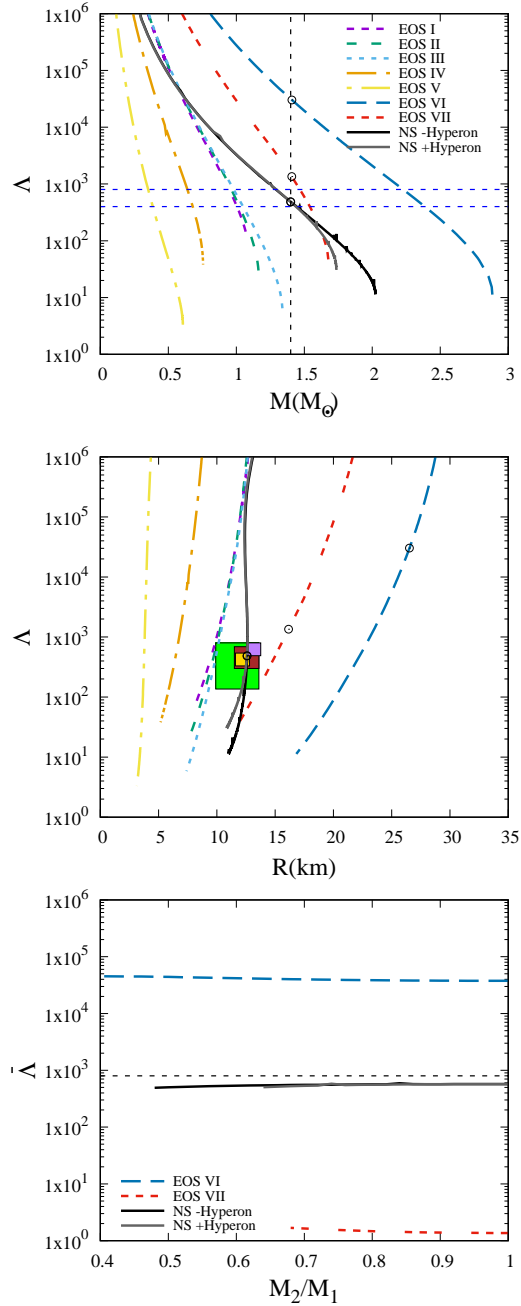}
	\caption{ The dimensionless tidal deformability $\Lambda$ vs the mass of fermionic DS. Vertical dashed line is $\Lambda$ for NS canonical mass ($M=1.4 M_{\odot}$) (Upper panel). Horizontal line is $\Lambda$ = 800 and 400 which are estimated from GW170817.  Plot of $\Lambda$ vs DS radius (Middle panel). Black circle is belong to $M=1.4 M_{\odot}$. The shaded rectangle are constraints from: Green \cite{21}, purple \cite{22}, brown \cite{23} and yellow \cite{24}. See detailed of the physics of the constraints in Ref. \cite{25}.  Plot of mass weighted tidal deformability in the binary of two compact DSs (lower panel). The chirp mass for the corresponding DS binary is fixed as the one of NS i.e., $M_{chirp}=1.188 M_{\odot}$. We can compare those $\bar{\Lambda}$ to that observed by LIGO/Virgo for NSs binary\cite{26}. Dashed horizontal line is $\bar{\Lambda}=800$, which is the upper limit estimated from GW170817.}
	\label{fig:8}
\end{figure}
\begin{figure}
	\centering
	\includegraphics[width=0.8\textwidth]{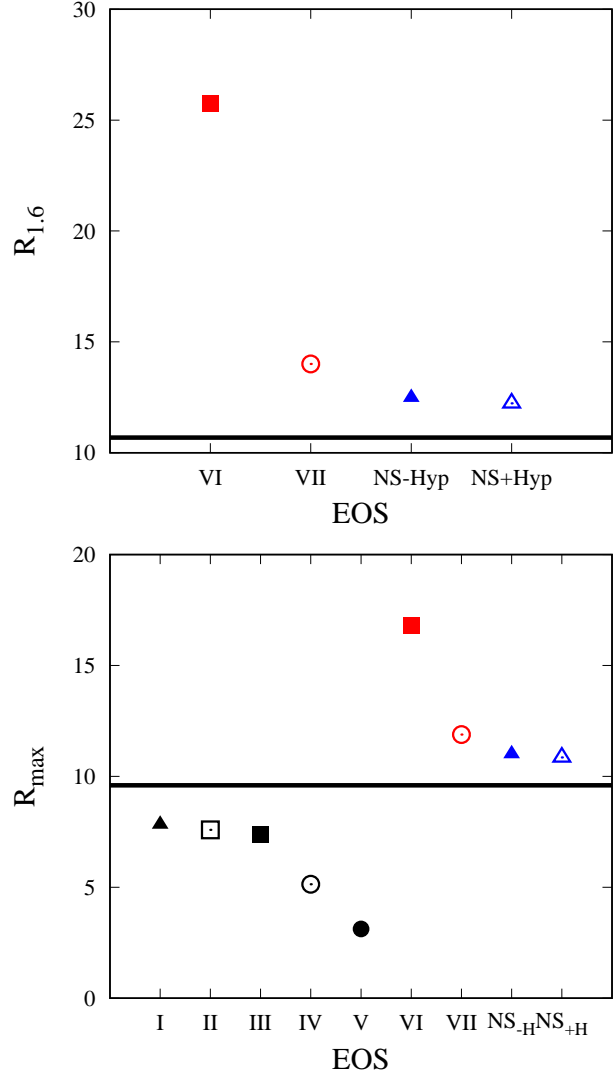}
	\caption{The radius $R_{1.6}$ (upper panel) and $R_{max}$ (lower panel) for non rotating DSs and NSs  of mass 1.6 $M_{\odot}$ and the maximum one. For NSs, the EOSs are calculated by using  RMF models. Black line in upper panel is $R_{1.6}$=$10.68_{-0.04}^{+0.14}$ km i.e., the radius constraint for NS with mass 1.6 $M_{\odot}$ \cite{27}. Black line in lower panel is radius constraint of NS maximum mass  i.e., $R_{max}$ =$9.60^{+0.14}_{-0.03}$ km \cite{27}.}
	\label{fig:9}
\end{figure}
\begin{figure}
	\centering
	\includegraphics[width=0.7\textwidth]{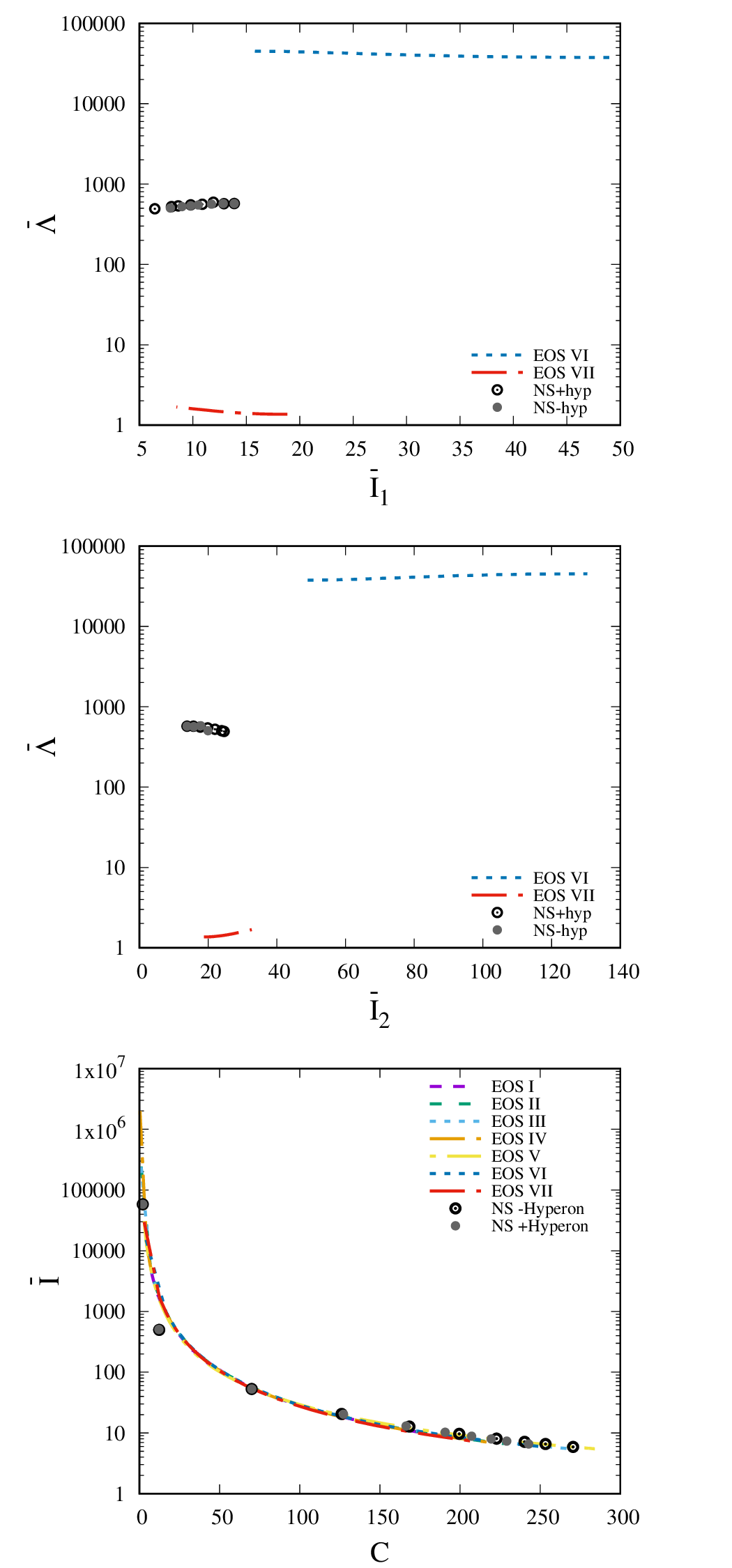}
	\caption{ Mass weighted average tidal deformability parameter $\bar{\Lambda}$ as the function of dimensionless moment of inertia of the lighter component of binary fermion DS with $M_{chrip}=1.188 M_{\odot}$ (Upper panel).  Same as the one in upper panel but for heavier component of binary fermion DS (Middle panel). Dimensionless moment of inertia as function of compactness (Lower panel). The ones of NSs are also given for comparison. }
	\label{fig:10}
\end{figure}

\subsection{Possible observations}
\label{subsec_observ}
In this subsection, we calculate the dimensionless tidal deformability of binary DSs as the main observable which can be conected to GW, radius constraint using the procedure from Bauswein {\it et al.}~\cite{27} as well as mass-weighted average tidal deformability parameter as a function of inertia moment and compactness of each star. The corresponding results will be compared to those of NSs and the corresponding constraints that extracted from GW170817. 

Note that the dimensionless tidal-deformability which can be related to DSs compactness which is defined as
\begin{equation}
  \Lambda\equiv \frac{2k_2}{3C^5}.
  \label{D_{less}TD}
\end{equation}
Note that  $ \Lambda$ is equal to $ \bar{\lambda}$ used in Fig.~\ref{fig:7}. In principle $ \Lambda$ can be deduced from the GW signals. However, how to distingush the GW signal from dark stars to those of other compact objects is another matter. Therefore, similar to that in Ref.~\cite{Sennett2017}, we hope, this study can shade a light for the question how to detact DSs using GW information in the future.

In binary objects, mass weighted tidal deformability $\bar{ \Lambda}$ is defined as
\begin{equation}
 \bar{ \Lambda} \equiv \frac{16}{13}\frac{(M_1+12 M_2)M_1^4 \Lambda_1+(M_2+12 M_1)M_2^4 \Lambda_2}{{(M_1+M_2)}^5},
  \label{BinaryD_{less}TD}
\end{equation}
where $M_1$, $M_2$ and $\Lambda_1$, $\Lambda_2$ are masses and dimensionless tidal deformabilites of two binary compact objects. Note that $\bar{ \Lambda}$=$\Lambda_1$=$\Lambda_2$ if  $M_1$= $M_2$. It can be seen in upper and middle panels of Fig. \ref{fig:8} that $\Lambda$s predicted by DSs studied here are quite sensitive to the EOS used. It can be observed in upper panel, only the ones of EOS VI, EOS VII are crossed NS canonical mass line and other EOSs cross the constraint from GW170817 for NS canonical mass i.e. 400 $<\Lambda<$ 800 earlier i.e., when M $<$ 1.4 $M_\odot$. Note that as it is expected that NSs EOS compatible with constraint from GW170817 for NS canonical mass. In middle panel we provide model dependent EOS of NS constraints of NS  $\Lambda$ vs radius obtained from Refs.~\cite{21,22,23,24} which are also shown and discussed in Ref.~\cite{25}. It is obvious that the corresponding  $\Lambda$ vs radius relations for DSs are beyond these NS constraints. In lower panels of Fig. \ref{fig:8}, the  mass weighted tidal deformability in the binary of two compact DSs predicted by EOS VI and VII as a function of $M2/M1$ using $M_{chirp}=1.188 M_{\odot}$ are shown. The result of NS and constraint  $\bar{ \Lambda}$=800 are also shown. This indicates that the $M_{chirp}$ of DSs binary with different traetment of boson scalar-vector coupling combinations with the same mass of DM particle, i.e., 1 GeV might be different and they might be also different compared to that of NSs. Therefore, the determination of $M_{chirp}$ of DSs binary is important. To this end, from the the plots shown in all panels of Fig. \ref{fig:8}, we can conclude that dimensionless  tidal deformability  of DSs can be distinguished to those of NSs. The difference might be observed from the future measurements of GW  events from DSs binary.

Recently, it is reported that the authors of Ref. \cite{27} have found a new and powerful method to constrain the radius of NS. They have shown that the total mass of GW170817 provides a reliable constraint on the stellar radius if the binary merger of  NS did not result in a prompt collapse as suggested by the interpretation of the associated electromagnetic emission. They claimed that these constraints are robust because they only need the chirp mass and a distinction between prompt and delayed collapse of the merger remnant, which may be inferred from the electromagnetic signal for NSs or from the presence/absence of ring-down GW signal (see details in Ref.~\cite{27} and the references therein). They have found that the radius constraint of NS as $R_{1.6}$ = $10.68_{-0.04}^{+0.14}$ km and $R_{max}$ = $9.60^{+0.14}_{-0.03}$ km. If the method  of \cite{27} could be also applied  for DSs binary merger and we have known already the corresponding  chirp mass, in principle, we could also constrain the radius of DSs. In Fig. \ref{fig:9}, we have shown the plots of radius $R_{1.6}$ in the upper panel and $R_{max}$ in lower panel for non-rotating DSs and NSs of mass 1.6 $M_{\odot}$ and the one of maximum mass. For NSs, the EOSs are calculated by using  RMF models with and without hyperons using BSP parameter set. It can be seen, for NS case that the radii are pretty close to those radii constraints of Ref. \cite{27}. However, for DS case, both radius predictions are away form NS constraint and they are quite sensitive to the stiffness of the  EOS. It means that the total mass of the DSs binary could be larger or smaller than the total mass of GW170817. If the total mass of the DSs binary can be measured in the future from GW signal measurements, then the mass and radius of DS could be determined.

For completeness, in upper panel of Fig.~\ref{fig:10}, we show the mass-weighted average tidal deformability parameter $\bar{\Lambda}$ as the function of dimensionless moment of inertia of the lighter component of binary DS $\bar{I_1}$ with $M_{chrip}=1.188 M_{\odot}$. While in the middle panel we show dimensionless moment of inertia of heavier component of binary DS $\bar{I_1}$. In the lower panel, we show the corresponding DSs dimensionless moment of inertia as a function of compactness. For comparison, we add also the corresponding results of NSs. From the lower panel, it can be understood that  I-Love universal relation for DSs is similar to that of NSs, because the dimensionless moment of inertia as the function of compactness relation of DSs are similar to that of NS, especially for large compactness value. However, the  $\bar{\Lambda}$ vs  $\bar{I_1}$ and  $\bar{\Lambda}$ vs  $\bar{I_2}$ depend sensitively on the stiffness of the EOS used. Therefore, in this way,  moment of inertia of DSs could be also distinguished from those of NSs indirectly from GW information.

To this end, we need to recap what we have found. If GW signal from DS binary can detect, it is closer to the one of NS than those other compact objects such as BH (significant different in compactness) and quark stars (different gravitational echo behavior). Note that Advanced LIGO  and Advanced Virgo gravitation wave facilities detect the neutron star binary in-spiral signal known as GW170817 signal. Ones of the important quantity that they measured from  GW170817 signals for low spin prior are the chirp mass $M_{chirp}$=1.188 $M_{\odot}$, combined dimensionless tidal deformability  $\bar{\Lambda}$ and  dimensionless tidal deformability for canonical mass  $\Lambda (1.4 M_{\odot})$, both $\le$ 800~\cite{26}. In this work, we have shown that DSs can have $M_{chirp}$ smaller or larger than 1.188 $M_{\odot}$ as well as the maximum  $\bar{\Lambda}$ or $\Lambda (1.4 M_{\odot})$ can be smaller or larger than 800, all depending on the stiffness of the corresponding EOS.  If any GW signal  can be measured with these criteria in the future, it could be indicated as the signals from DS binary. Then the mass of DM particle, the coupling of dark particle with scalar and vector bosons might be deduced from the corresponding tidal deformability information  of the DS binary. Note that these study can be performed also for other models of interacting dark particle including bosonic and fermionic types. We believe for other models of interacting dark particle that qualitatively the results could be similar but the details will be different.  

\section{CONCLUSIONS}
\label{sec_conclu}

We have investigated the role of dark particles mass and scalar boson exchange as a mediator of the interaction on compact stars consisting fermionic DM by studying the bulk properties of slow rotating DSs including their moment of inertia and tidal deformability. We have found that the role of attractive scalar boson exchange and the mass of DM particles to control the stiffness of the EOS of DSs are crucial. If we increase the strength of scalar boson contribution then the DSs could be more compact. Therefore, if scalar boson exchange contribution is included, the DSs compactness could exceed $C$=0.22. This shows that $C$=0.22 is not the genuine property of DSs. This compactness upper bound could not be used to distinguish between DSs and NSs. If we compare the DSs moment of inertia and tidal deformation to those of NSs (with and without hyperons) predicted by relativistic mean field model with BSP parameter set, it is obvious that the features of both stars are quite different. The difference in the moment of inertia can detect indirectly from tidal deformability information due to the fact that the universal relation between the moment of inertia and tidal deformation of compact objects exists. We also found that the universal I-Love relation is not affected by scalar boson exchange contribution and DM particles mass. We have also discussed the possible observations which can be extracted from the information of the corresponding gravitation wave signals such as compactness, gravitational wave echo, tidal deformability, and radii constraints. The compactness can be used to distinguish DSs and black holes. The gravitational wave echo could distinguish the DSs with strange stars by observing whether the mass-radius plot crosses the photon-sphere line, while tidal deformability and radii constraint could distinguish the DSs with NSs.  The authors of Ref.~\cite{Sennett2017} have already studied systematically how to distinguish boson stars from black holes and NSs from tidal interaction in in-spiraling systems. We believe that this analysis can be used also to distinguish DSs from boson stars and black holes.  However, such work is already outside of this present work.  The mass of DM particle, the coupling of dark particle with scalar and vector bosons might be also deduced from future measurements of  GW signal of DS binary.

\section*{ACKNOWLEDGMENT}
This research is funded by PITTA grant No.~2230/UN2.R3.1/HKP.05.00/2018. A.B.W. would like also to thanks to Prof. T. Mart for the discussions in the early stage of this work.


\end{document}